\documentclass[prl,twocolumn,reprint]{revtex4-2}

\usepackage{xcolor}
\usepackage{amssymb}
\usepackage{amsmath}
\usepackage{graphicx}
\usepackage{float}
\usepackage{braket}
\usepackage{bbm}
\usepackage{color}
\usepackage{comment}
\usepackage{gensymb}
\usepackage{gensymb}
\usepackage{lipsum}
\usepackage{braket}
\usepackage{todonotes}
\definecolor{blue}{rgb}{0,0,1}
\definecolor{grey}{rgb}{0.6,0.6,0.6}

\usepackage[colorlinks=true,citecolor=blue,linkcolor=blue,urlcolor=blue]{hyperref}

\begin{document}
	
	\title{Autonomous Distribution of Programmable Multiqubit Entanglement in a Dual-Rail Quantum Network}
	
	\author{J. Agustí$^{1,2,3}$, X. H. H. Zhang$^{1,2,3}$, Y. Minoguchi,$^4$,  P. Rabl$^{1,2,3,4}$}
	\affiliation{$^1$Technical University of Munich, TUM School of Natural Sciences, Physics Department, 85748 Garching, Germany} 
	\affiliation{$^2$Walther-Meißner-Institut, Bayerische Akademie der Wissenschaften, 85748 Garching, Germany}
	\affiliation{$^3$Munich Center for Quantum Science and Technology (MCQST), 80799 Munich, Germany} 
	\affiliation{$^4$Vienna Center for Quantum Science and Technology, Atominstitut, TU Wien, 1020 Vienna, Austria}

	\date{\today}
	
	\begin{abstract}
		We propose and analyze a scalable and fully autonomous scheme for preparing spatially distributed multiqubit entangled states in a dual-rail waveguide QED setup. In this approach, arrays of qubits located along two separated waveguides are illuminated by correlated photons from the output of a nondegenerate parametric amplifier. These photons drive the qubits into different classes of pure entangled steady states, for which the degree of multipartite entanglement can be conveniently adjusted by the chosen pattern of local qubit-photon detunings. Numerical simulations for moderate-sized networks show that the preparation time for these complex multiqubit states increases at most linearly with the system size and that one may benefit from an additional speedup in the limit of a large amplifier bandwidth. Therefore, this scheme offers an intriguing new route for distributing ready-to-use multipartite entangled states across large quantum networks, without requiring any precise pulse control and relying on a single Gaussian entanglement source only.
	\end{abstract}
	
	\maketitle
	
	\emph{Introduction.}---As quantum computing and quantum communication systems with an increasing number of coherently integrated components become technologically available, a growing demand for efficient schemes to transfer quantum states or distribute entanglement across different parts of such networks will arise~\cite{Cirac97,Northup14,Reiserer15,Awschalom21}. While basic protocols to do so are well known and have already been successfully implemented in a variety of platforms~\cite{Moehring07,Ritter12,Hensen15,Riedinger18,Kurpiers18,Bienfait19,Fedorov2021,Pompili23,Krutyanskiy23,Sahu23}, it is envisioned that in future quantum devices, entanglement must be generated and interchanged among many thousands of qubits within a limited coherence time. In view of this challenge, there is a strong motivation to go beyond a serial application of existing protocols and search for more efficient quantum communication strategies that are fast, parallelizable, and, ideally, require a minimal amount of classical control.

\begin{figure}
		\centering		
		\includegraphics[width=\columnwidth]{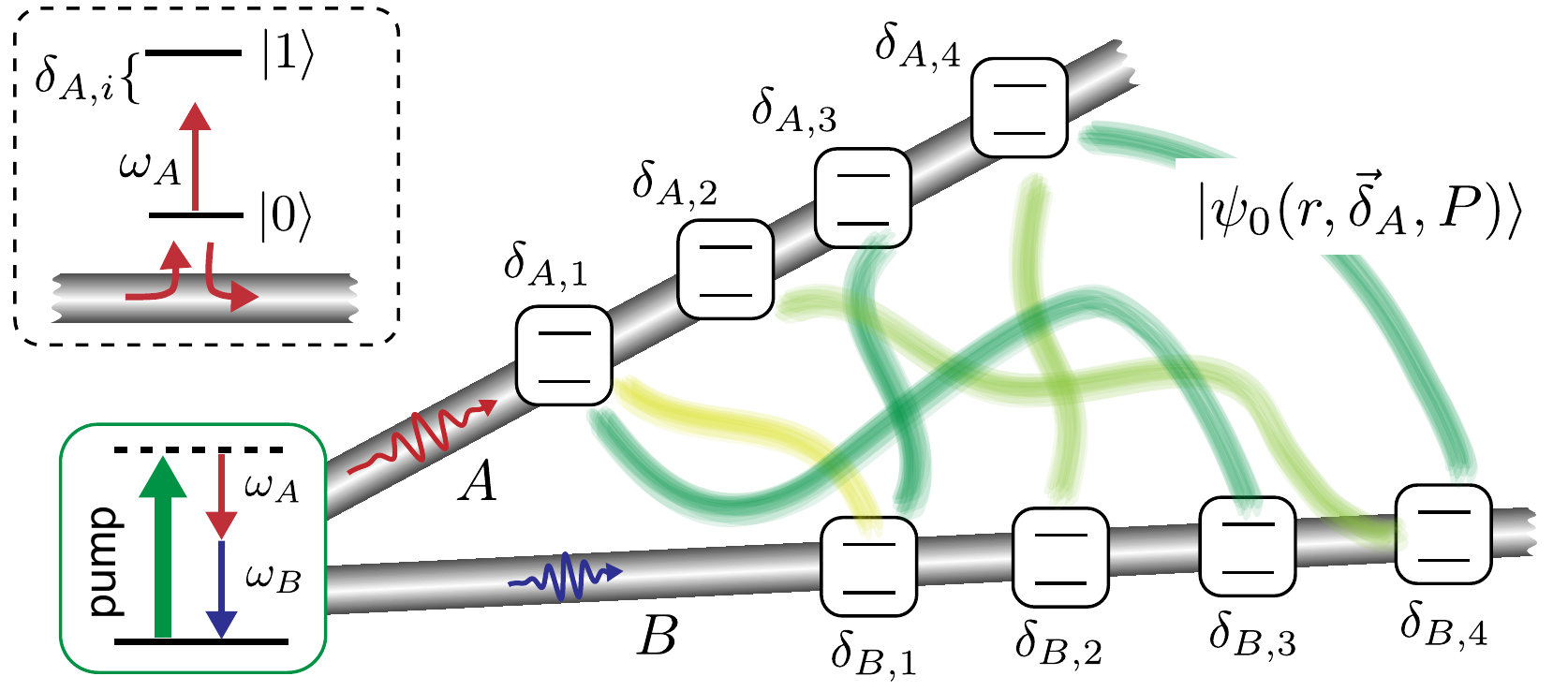}
		\caption{Sketch of a dual-rail quantum network, where qubits along two separated waveguides are driven by the correlated output of a nondegenerate parametric amplifier and relax into a pure steady state $|\psi_0(r,\vec \delta_A,P)\rangle$. As shown in the inset, the qubits in waveguide $A$ ($B$) are detuned from the central photon frequency $\omega_A$ ($\omega_B$) by $\delta_{A,i}$ ($\delta_{B,i}$) and the qubit-waveguide coupling is assumed to be fully directional. See text for more details. }
		\label{fig:Setup}
	\end{figure}

	In this Letter, we describe a fully autonomous entanglement distribution scheme, which exploits an intriguing physical effect, namely the formation of multipartite entangled stationary states in a cascaded dual-rail quantum network. Specifically, we consider a configuration as shown in Fig.~\ref{fig:Setup}, where spatially separated qubits located along two photonic waveguides are illuminated by the correlated output of a nondegenerate parametric amplifier~\cite{Carmichael99}.  Previously, it was proposed to use broadband squeezed reservoirs for generating bipartite entanglement between separated qubit pairs~\cite{Kraus04,Paternostro04,Blais17,Mirrahimi18,You18,Govia22,Agusti22} or, for specific arrangements, between qubits along a 1D channel~\cite{Agarwal90,GutierrezJauregui23} or in coupled arrays~\cite{Zippilli13,Clerk22}.  Here we show, first of all, that this concept can be generalized to produce, under ideal conditions, an arbitrary number of maximally entangled qubit pairs over large distances. Moreover, we find that the entanglement shared between different sets of qubits can be adjusted by simply changing the local qubit-photon detunings. This provides a convenient way to “program” different classes of multipartite entangled states without the need for any time-dependent control or additional nonlocal operations.

	To evaluate the scalability of this approach, we simulate the formation of these multipartite entangled states under more realistic conditions, taking in particular a finite bandwidth of the squeezing source into account. We find that the maximal number of entangled qubit pairs, $N_{\rm ent}$, remains rather robust under the influence of experimental imperfections and that the total preparation time, $T_{\rm prep}\sim N_{\rm ent}$, scales at most linearly with the system size, independently of the complexity of the prepared state. In the limit of a large amplifier bandwidth, the intrinsic parallelization of the preparation scheme can be exploited to further reduce $T_{\rm prep}$,  which shifts the technological requirements for scalability from the control of many qubits to the optimization of a single Gaussian squeezing source. This can be advantageous for many applications in optical, microwave, or hybrid~\cite{Xiang13,Kurizki15,Clerk20,Han21} quantum networks, where such photonic devices are currently developed~\cite{Walmsley11,Couteau18,Eichler11,Menzel12,Huard12,Winkel20,Peugeot21,Sahu23}.

	\emph{Model.}---We consider a dual-rail quantum network as depicted in Fig.~\ref{fig:Setup}, where two sets of qubits $\eta=A,B$ are coupled to two separate photonic channels. The waveguides are connected to a common nondegenerate parametric amplifier, which we model by a two-mode squeezing interaction $(\hbar=1)$ $H_\chi=ig (a_A^\dagger a_B^\dagger-a_Aa_B)$ for two local modes with bosonic annihilation operators $a_A$ and $a_B$. These photons then decay into the respective waveguides with rate $\kappa$ and drive the qubits into a correlated state. For the following analysis, we assume that the qubit-waveguide coupling is fully directional~\cite{Stannigel12,Pichler15,Lodahl17} and label the qubits by the index $i=1,\dots,N$ along the direction of propagation. Such conditions can be realized by using circulators~\cite{Sliwa15,Kerckoff15,Chapman17,Lecocq17,Masuda19,Wang21}, chiral waveguides~\cite{Lodahl17}, or other schemes for directional coupling~\cite{Guimond20,Gheeraert20,Kannan22,Joshi2023}.

	We first focus on the limit of a broadband amplifier, $\kappa \rightarrow \infty$, in which case the dynamics of the photons can be adiabatically eliminated to obtain an effective master equation (see~\cite{Supp} for more details) 
	\begin{equation}
		\label{Eq:ME}
		\dot \rho_{\rm q}= -i[H_{\rm casc},\rho_{\rm q}]+  \sum_{\eta=A,B} \gamma \mathcal{D}[J_\eta]\rho_{\rm q}
	\end{equation}
	for the reduced qubit density operator $\rho_{\rm q}$. Here $\gamma$ denotes the decay rate of each individual qubit and $\mathcal{D}[C]\rho=C\rho C^\dagger-\{C^\dag C, \rho\}_+/2$. In Eq.~\eqref{Eq:ME} we have already rewritten the underlying directional qubit-qubit interactions in terms of a coherent Hamiltonian evolution with 
	\begin{equation}\label{eq:Hcasc}
		H_{\rm casc}=\sum_{\eta,i}\frac{\delta_{\eta,i}}{2}\sigma_{\eta,i}^{z}+i\frac{\gamma}{2}\sum_{\eta,j>i}\left( \sigma^+_{\eta,i}\sigma_{\eta,j}^--{\rm H.c.} \right),
	\end{equation}
	and purely dissipative processes with collective jump operators
	\begin{flalign}
		\label{Eq:Dissipators}
		&J_A=\cosh(r) L_A-\sinh(r)L_B^\dagger,\\
		&J_B=\cosh(r)L_B-\sinh(r)L_A^\dagger,
	\end{flalign}
	where $L_{\eta}=\sum_{i=1}^{N}\sigma_{\eta,i}^-$. In this broadband limit, the system is thus fully determined by the squeezing parameter $r = 2 \tanh^{-1}(2g/\kappa)$, characterizing the degree of two-mode squeezing of the photon source, and the two sets of qubit detunings, $\vec \delta_{\eta=A,B}=(\delta_{\eta,1}, \delta_{\eta,2},\dots, \delta_{\eta,N})$.    
	
	\emph{Steady states.}---Equation~\eqref{Eq:ME} describes an open quantum many-body system with competing coherent and dissipative processes, which in general drive the qubits into a highly mixed and not very useful steady state. However, in the following, we show that there exist specific conditions under which the steady state of the network, $\rho_{\rm q}^0   =|\psi_0\rangle \langle \psi_0|$, is not only pure but also exhibits different degrees of multipartite entanglement that can be controlled by the local detunings $\delta_{\eta,i}$.   
	
	We start our analysis by considering the simplest case of a single pair of qubits ($N=1$) and $\delta_{A,1}=\delta_{B,1}=0$, as originally discussed in Ref.~\cite{Kraus04}. In this case, one can explicitly show that the unique steady state of Eq.~\eqref{Eq:ME} is  $|\psi_0\rangle = |\Phi^+_{1,1}\rangle$, where
	\begin{equation}
		\label{Eq:BellPair}
		|\Phi^+_{i,j}\rangle=\frac{\cosh(r)|0_{A,i}\rangle|0_{B,j}\rangle+\sinh(r) |1_{A,i}\rangle|1_{B,j}\rangle}{\sqrt{\cosh(2r)}}
	\end{equation}
	approaches a maximally entangled Bell state for $r\gg1$. This state satisfies the dark-state conditions $J_\eta |\psi_0\rangle  =0$ and $H_{\rm casc} |\psi_0\rangle= 0$, which implies that once the qubits have reached the steady state, they completely decouple from the squeezed photonic bath. Consequently, they no longer affect successive qubits along the waveguide. 
	
	Importantly, this observation remains true even for finite detunings satisfying $\delta_{A,1}+ \delta_{B,1}=0$, which then allows us to systematically identify also more complex multiqubit steady states by proceeding in two steps. First, we set  $\vec \delta_B=-\vec \delta_A$, such that, according to the argument from above, qubits with the same index decouple pairwise from the photonic reservoir. The network then  relaxes into the pure steady state $|\psi_0\rangle = |\Phi_\parallel\rangle$, where  
	\begin{equation}
		\label{Eq:GeneralDarkState}
		|\Phi_\parallel\rangle= \bigotimes_{i=1}^N |\Phi^+_{i,i}\rangle
	\end{equation}
	is the product of $N$ consecutive Bell pairs of the type given in Eq.~\eqref{Eq:BellPair}. Interestingly, this result is independent of the total number of qubit pairs.

	In the second step, we make use of the form invariance of the cascaded master equation in Eq.~\eqref{Eq:ME} under unitary transformations of the type~\cite{Stannigel12} 
	\begin{equation} 
		\label{eq:Unitary}
		U_{i,i+1}= e^{i\theta_{i,i+1} (\vec{s}_{B,i}+ \vec{s}_{B,i+1})^2},
	\end{equation}
	where $\vec s_{\mu}=(\sigma_\mu^x,\sigma_\mu^y,\sigma_\mu^z)/2$ and the mixing angle satisfies $\tan(\theta_{i,i+1})=(\delta_{B,i}-\delta_{B,i+1})/\gamma$. Under these transformations, one finds that  $U_{i,i+1}J_\eta U^\dag_{i,i+1}= J_\eta$ and 
	\begin{eqnarray}\label{eq:FormInvariance}
		U_{i,i+1} H_{\rm casc}(\vec \delta_A, \vec \delta_B) U^\dag_{i,i+1}&=& H_{\rm casc}(\vec \delta_A, P_{i,i+1}\vec \delta_B),
	\end{eqnarray} 
	where the permutation $P_{i,i+1}$ exchanges $\delta_{B,i}$ and $\delta_{B,i+1}$. In other words, given a pure steady state $|\psi_0\rangle$ for a certain detuning pattern $\vec \delta_B$, the state $|\psi_0^\prime\rangle=U_{i,i+1}|\psi_0\rangle$ is a pure steady state of the same network with a permuted pattern of detunings, $\vec \delta_B^\prime= P_{i,i+1}\vec \delta_B$. 
	
\begin{figure}
		\centering		
		\includegraphics[width=\columnwidth]{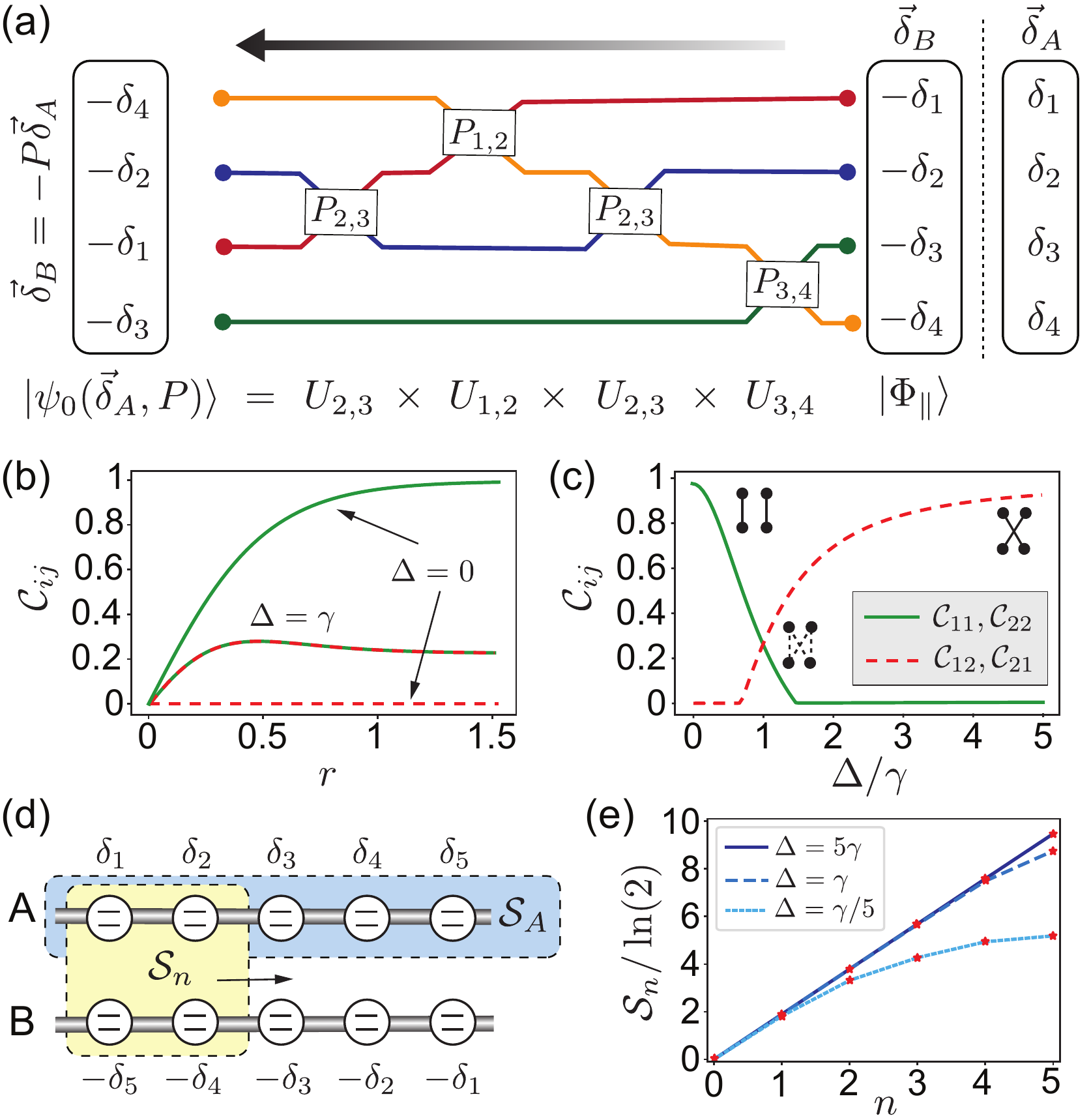}
		\caption{(a) Graphical illustration of Eq.~\eqref{eq:SteadyStateGeneral}. Starting from $\vec \delta_B=-\vec \delta_A$, the detunings in waveguide B are reordered as $\vec \delta_B=-P\vec \delta_A$ through nearest-neighbor transpositions, following the colored lines as a guide to the eye. Each transposition maps into one of the unitary operations $U_{i,i+1}$ that determine the final steady state.  (b) Bipartite entanglement expressed in terms of the concurrences $\mathcal{C}_{ij}$ for the four-qubit state in Eq.~\eqref{eq:fourqubit} as a function of $r$, and in (c) as a function of $\Delta$ for $r=1$. (d) Sketch of the detuning pattern for the family of multipartite states described in the text and different partitions for evaluating the entanglement entropy. (e) Entanglement entropy $\mathcal{S}_n$ as a function of $n$, for different detunings $\Delta$ and $r=1$.}
		\label{fig2:Entanglement}
	\end{figure}

	This form invariance now allows us to construct a large family of multipartite entangled steady states, which are parametrized by (i) the squeezing parameter $r$, (ii) the set of detunings $\vec \delta_A$ for qubits in waveguide $A$ and (iii) a permutation $P$ that fixes the detunings in waveguide $B$ to be $\vec \delta_B = - P\vec \delta_A$. By decomposing $P= \prod_\sigma P_{i_\sigma,i_\sigma+1}$ into a product of nearest-neighbor transpositions, we can start with the state in Eq.~\eqref{Eq:GeneralDarkState} and then use the relation below Eq.~\eqref{eq:FormInvariance} to derive an explicit expression for the corresponding steady state,  
	\begin{equation}\label{eq:SteadyStateGeneral}
		|\psi_0(r,\vec \delta_A,P)\rangle = \prod_\sigma U_{i_\sigma,i_\sigma+1}  |\Phi_\parallel\rangle.
	\end{equation}
	Importantly, this is also the unique steady state of the network, as discussed in more detail in~\cite{Supp}. A graphical illustration of Eq.~\eqref{eq:SteadyStateGeneral} is presented in Fig.~\ref{fig2:Entanglement} (a).

	\emph{Entanglement.}---To investigate the entanglement properties of the family of states in Eq.~\eqref{eq:SteadyStateGeneral}, we start with the case $N=2$ and choose the only nontrivial permutation $P=P_{1,2}$. We obtain
	\begin{equation}\label{eq:fourqubit}
		\ket{\psi_0}=\frac{\gamma |\Phi^+_{1,1}\rangle|\Phi^+_{2,2}\rangle+i\Delta|\Phi^+_{1,2}\rangle|\Phi^+_{2,1}\rangle}{\sqrt{\gamma^2+\Delta^2}},
	\end{equation}
	where  $\Delta=\delta_{A,1}-\delta_{A,2}$.  In Figs.~\ref{fig2:Entanglement} (b)  and (c) we visualize the entanglement structure of this state in terms of the concurrences $\mathcal{C}_{ij}\equiv \mathcal{C}(\rho_{A,i|B,j})$~\cite{Hill97,Horodecki09} of the reduced bipartite qubit states, $\rho_{A,i|B,j}$. For $\Delta=0$, we find that for parallel pairs $\mathcal{C}_{ii}\simeq 1$ already for moderate values of $r\gtrsim 1$, consistent with the state $|\Phi_\parallel\rangle$. For $|\Delta| \gg\gamma $ the same is true for diagonal pairs, i.e., $\mathcal{C}_{12}=\mathcal{C}_{21}\simeq 1$. For all intermediate parameters, the state is a genuine four-partite entangled state~\cite{Guhne11}, and belongs to the set of locally maximally entanglable states~\cite{Kruszynska09} for $r\gg1$.
	
	For a larger number of qubits, we can use the entanglement entropy $\mathcal{S}(\rho_r)=-{\rm Tr}\{\rho_r \ln \rho_r\}$ for a reduced state $\rho_r$ to study the entanglement between different bipartitions of the network. First of all, this analysis shows that 
	$\mathcal{S}_A\equiv\mathcal{S}(\rho_{A}) = -N \ln\left[x^x(1-x)^{(1-x)}\right]$, where $x=\cosh^2(r)/\cosh(2r)$,
	only depends on the squeezing parameter $r$. This can be understood from the fact that the unitaries $U_{i,i+1}$ only act within subsystem $B$. Thus, with respect to this partition, the states in Eq.~\eqref{eq:SteadyStateGeneral} can be understood as generalized “rainbow states”~\cite{Vitagliano10,Zhang17,Clerk22} with a volume-law entanglement $\mathcal{S}(\rho_{A}) \simeq N\ln 2$ for $r\gtrsim 1$. In contrast, for partitions along the chain, the entanglement entropy $\mathcal{S}_n=\mathcal{S}(\rho_{[1,\dots,n]})$ depends not only on the chosen permutation $P$, but also on the pattern of detunings $\vec \delta_A$. This is illustrated in Figs.~\ref{fig2:Entanglement} (d) and (e), where we consider as an example the detunings $\delta_{A,i}=(i-1)\Delta$ and the reversed order, $\delta_{B,i}=-P_{\rm rev}\delta_{A,i}=-\delta_{A,N+1-i}$,  in waveguide $B$. For $\Delta \gg \gamma$ the unitaries in  Eq.~\eqref{eq:SteadyStateGeneral} correspond to approximate SWAP operations and $\mathcal{S}_n\simeq  2 n \ln 2$. Instead, for $\Delta\lesssim \gamma$, the entangling unitaries $U_{i,i+1}\approx \sqrt{\rm SWAP}$ generate more multipartite entanglement across the whole chain, which reduces the block-entanglement $\mathcal{S}_n$ correspondingly.  In general, different choices for $\vec \delta_A$ and $P$ can be used to define certain blocks of qubits that are entangled among each other, independently of their physical location.

\begin{figure}
		\centering		
		\includegraphics[width=\columnwidth]{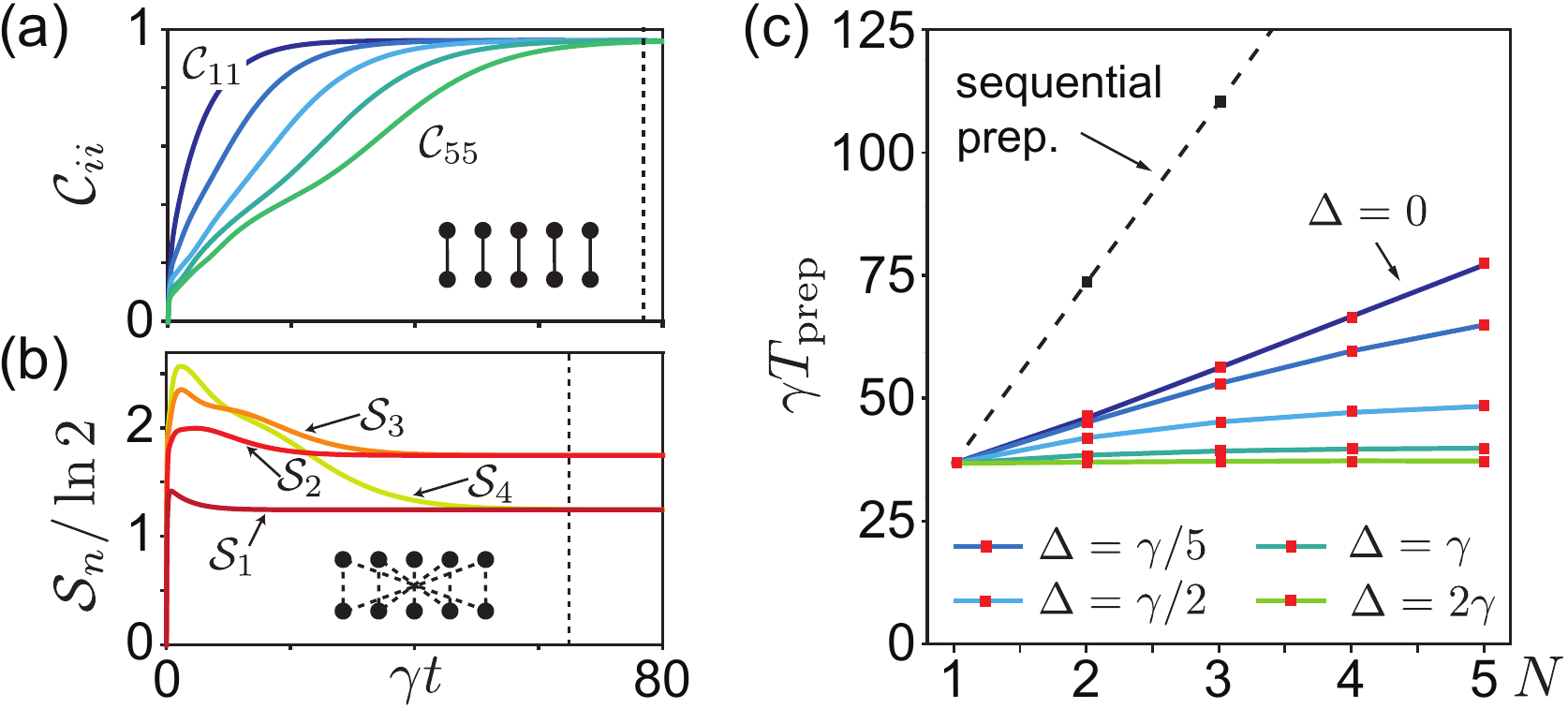}
		\caption{(a) Relaxation into a bipartite entangled state for $\vec \delta_A=0$ and (b) into a multipartite entangled state for $\vec \delta_B =-P_{\rm rev} \vec \delta_A$ and $\Delta=\gamma/5$. In both cases $N=5$.
			(c) Scaling of the preparation time $T_{\rm prep}$ for different ratios $\Delta/\gamma$, where $\delta_{A,i}=\Delta (i-1)$ and $\vec \delta_B =-\vec \delta_A$. 
			We define $T_{\rm prep}$ via the condition $[1-\mu(T_{\rm prep})]/N= 0.001$, where $\mu={\rm Tr}[\rho_{\rm q}^2]$ is the purity. For the examples in (a) and (b), $T_{\rm prep}$ is indicated by the dashed vertical line. In all plots $r=1$.}
		\label{fig:Fig3Time}
	\end{figure}

	\emph{Preparation time.}---So far we have shown that a single two-mode squeezing source is in principle enough to entangle an arbitrary number of qubits. However, for practical applications, we must still evaluate the time $T_{\rm prep}$ that it takes to prepare this state. To do so we first continue with the analysis of the ideal qubit master equation in Eq.~\eqref{Eq:ME} and study the relaxation dynamics toward the steady state $|\psi_0\rangle$, assuming that at $t=0$ all qubits are initialized in state $|0\rangle$. In Fig.~\ref{fig:Fig3Time} this evolution is shown in (a) for the bi-partite entangled state $|\Phi_\parallel\rangle$ with $\vec \delta_A=0$ and in (b) for the multipartite entangled state considered in Fig.~\ref{fig2:Entanglement} (e). 
	In the bipartite case, we observe a successive, pairwise formation of Bell states with a total time $T_{\rm prep}\sim N$. Interestingly, already for $\delta_{A,i}=0$, this preparation time is faster than a sequential preparation of $N$ independent Bell pairs, i.e., $T_{\rm prep}(N) < N T_{\rm prep}(N=1)$. For detuned qubits
	the preparation time decreases further and $T_{\rm prep}(N)\simeq T_{\rm prep}(N=1)$ for $\Delta\gtrsim \gamma$, i.e., all pairs are prepared in parallel. For multipartite entangled states, where the differences $|\delta_{A,i}-\delta_{A,j}|$ are necessarily small, a full parallelization is not possible, but even in this case we obtain an intrinsic advantage compare to a sequential distribution of entanglement, followed by local gates. Note that for the same detunings $\vec \delta_A$, the relaxation time $T_{\rm prep}$ is independent of the permutation $P$.

	\emph{Scalability.}---All the results so far have been derived within the infinite-bandwidth approximation, which underlies Eq.~\eqref{Eq:ME} and assumes that correlated photons are available at arbitrary detunings. Obviously, this assumption must break down when $\delta_{\rm max} = {\rm max} \{ |\delta_{A,i}|\}  \gtrsim \kappa$, but even for $\delta_{A,i}=0$ it has been shown that any finite $\kappa $ limits the transferable entanglement~\cite{Agusti22}. Therefore, to provide physically meaningful predictions about the scalability of the current scheme it is necessary to go beyond the assumption of a Markovian squeezed reservoir~\cite{Kraus04,Paternostro04,Blais17,Mirrahimi18,You18,Govia22,Agusti22,Agarwal90,GutierrezJauregui23,Zippilli13,Clerk22} and take finite-bandwidth effects into account. To do so we now simulate the dynamics of the state of the full network, $\rho$, as described by the cascaded quantum master equation~\cite{Supp}
	\begin{equation}\label{eq:FullME} 
		\begin{split}
			\dot \rho =& -i[H_\chi , \rho]+ \sum_{\eta} \kappa \mathcal{D}[a_\eta]\rho \\
			& \sum_{\eta,i} \left( -i \frac{\delta_{\eta,i}}{2} [\sigma_{\eta,i}^{z},\rho] + \gamma \mathcal{D}[\sigma_{\eta,i}^-] \rho  + \frac{\gamma_\phi}{2} \mathcal{D}[\sigma_{\eta,i}^z] \rho\right)  \\
			&+\sum_{\eta,i}\sqrt{\kappa\gamma} \mathcal{T}[a_\eta,\sigma_{\eta,i}^-]\rho +   \sum_{\eta,j> i}  \gamma \mathcal{T}[\sigma^-_{\eta,i},\sigma_{\eta,j}^-]\rho.
		\end{split}
	\end{equation}
	Here we have already included a finite dephasing rate $\gamma_\phi$ for each qubit and introduced the superoperator $\mathcal{T}[O_1,O_2]\rho = [O_1\rho,O_2^\dag ]+ [O_2 ,\rho O_1^\dag]$ to model directional interactions between all nodes along the same waveguide.

\begin{figure}
		\centering		
		\includegraphics[width=\columnwidth]{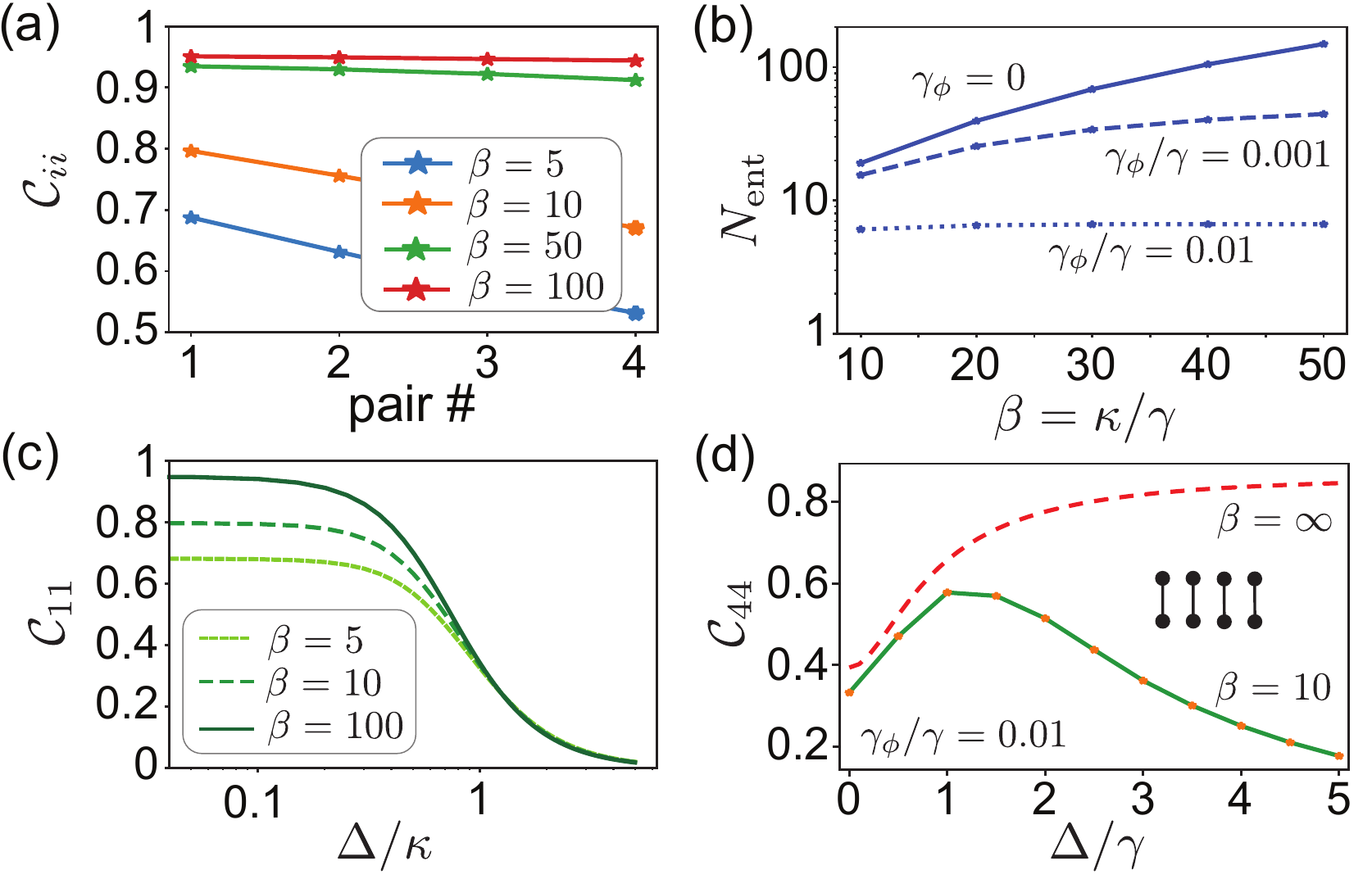}
		\caption{(a) Plot of the steady-state concurrences $\mathcal{C}_{ii}$ for $\vec \delta_A=0$ and different amplifier bandwidths. (b) Maximal number of entangled pairs, $N_{\rm ent}$ as a function of $\beta=\kappa/\gamma$ and different dephasing rates $\gamma_\phi$. (c) Dependence of the concurrence of a single qubit pair on the detuning $\Delta$, where $\delta_{A,1}=-\delta_{B,1}=\Delta$ and different values of $\beta$ have been assumed. (d) Plot of the concurrence $\mathcal{C}_{44}$ in a chain of $N=4$ qubit pairs with $\delta_{A,i}=(i-1)\Delta=-\delta_{B,i}$ and a finite dephasing rate. This plot illustrates the initial gain from a parallel preparation when $\Delta>0$, while the entanglement decreases again when $\delta_{\rm max}=(N-1)\Delta \approx \kappa$, due to finite bandwidth effects. In all plots $r=1$.}
		\label{fig:Fig4Scaling}
	\end{figure}

	In Fig.~\ref{fig:Fig4Scaling} (a) we plot the steady state concurrences $\mathcal{C}_{ii}$ for the case $\delta_{A,i}=0$ and different ratios $\beta=\kappa/\gamma$. We see that a finite bandwidth $\kappa$ reduces the maximal amount of entanglement for the first pair~\cite{Agusti22} and also results in a gradual decay of the entanglement along the chain. By using a linear extrapolation, $N_{\rm ent}=\mathcal{C}_{11}/(\mathcal{C}_{11}-\mathcal{C}_{22})$, we can use these finite-size simulations to extract the maximal number of pairs that can be entangled for a given $\beta$ and dephasing rate $\gamma_\phi$. These results are summarized in Fig.~\ref{fig:Fig4Scaling} (b). We see that for otherwise ideal conditions, rather large numbers of $N_{\rm ent}\sim 10-100$ can be entangled for moderate $\beta$, while the presence of dephasing or other imperfections sets additional limits on $N_{\rm ent}$. Note that these results are for $\delta_{A,i}=0$, where the formation of the steady state is the slowest. Thus, these results represent approximate upper bounds for $N_{\rm ent}$ also for all other classes of multipartite entangled states.  Additional plots for number of entangled pairs for various experimental sources of imperfections together with estimates for the achievable $N_{\rm ent}$ in state-of-the-art microwave networks are presented in~\cite{Supp}.

	Finally, let us return to the observed speedup for far-detuned qubits, but taking a finite amplifier bandwidth into account.  In Fig.~\ref{fig:Fig4Scaling} (c) we investigate, first of all, the dependence of $\mathcal{C}_{11}$ on the detuning $\delta_{A,1}=\Delta$. As expected, this plot shows a significant decay of the entanglement for $\Delta/\kappa>1$, from which we also deduce that $\delta_{\rm max}<\kappa$ must be satisfied in the multiqubit case. 
	Since for a parallel preparation with $T_{\rm prep}(N)\sim \text{const}$ we require $\delta_{\rm max}\approx \gamma N$, we conclude that the number of pairs that can be entangled in parallel, $N_{\parallel}\approx N_{\rm ent}$, is actually comparable to the total number of entangled pairs for $\vec \delta_A=0$.  As a minimal illustration of this behavior, we consider in Fig.~\ref{fig:Fig4Scaling} (d) the example of $N=4$ pairs with $\delta_{A,i}=\Delta (i-1)$. We plot the concurrence of the last pair, $\mathcal{C}_{44}$, for a fixed dephasing rate $\gamma_\phi$ and increasing detuning $\Delta$. Up to $\Delta\sim \kappa$, entanglement increases due to a reduced preparation time, while for larger detunings finite-bandwidth effects set in and degrade the entanglement again. Note that for a parametric amplifier with asymmetric decay rates, $\kappa_A\neq \kappa_B$, the structure of the ideal qubit master equation in Eq.~\eqref{Eq:ME} remains the same~\cite{Supp}, but finite-bandwidth effects are determined by the minimal rate $\kappa_{\rm min}={\rm min}\{\kappa_A,\kappa_B\}$.

	\emph{Conclusions.}---In summary, we have presented a fully autonomous scheme for distributing entanglement among two distant sets of qubits. Within the same setup, states with varying degrees of bi- and multipartite entanglement can be prepared by adjusting the squeezing strength and the local qubit detunings, while retaining a preparation time that scales at most linearly with $N$. Compared to related autonomous protocols discussed for single waveguides~\cite{Stannigel12,Pichler15,Agarwal90,GutierrezJauregui23}, locally coupled chains~\cite{Zippilli13,Clerk22}, or combinations thereof~\cite{Lingenfelter23}, the use of a propagating two-mode entangled source offers the possibility to entangle qubits that are arbitrarily far apart~\cite{Supp} and a systematic way to parallelize the scheme by increasing the bandwidth of the amplifier. This makes this approach very attractive for long-distance entanglement distribution schemes with long-lived spins or narrow-bandwidth optical emitters, but also for local area quantum networks~\cite{Magnard20,Cleland21,Niu23,Renger23}, where multiple nodes can be simultaneously entangled with a limited amount of control.

	We thank Aashish Clerk, Matthias Englbrecht, Tristan Kraft, Barbara Kraus, and Kirill Fedorov for many stimulating discussions.  This work was supported by the European Union's Horizon 2020 research and innovation program under Grant Agreement No. 899354 (SuperQuLAN) and the Deutsche Forschungsgemeinschaft (DFG, German Research Foundation) No. 522216022. Most of the computational results presented were obtained using the CLIP cluster \cite{WebCLIP}. This research is part of the Munich Quantum Valley, which is supported by the Bavarian state government with funds from the Hightech Agenda Bayern Plus.

	\newpage
	
	\widetext
	
	\clearpage
	
	\setcounter{equation}{0}
	\setcounter{figure}{0}
	\setcounter{table}{0}
	\setcounter{page}{1}
	\makeatletter
	\renewcommand{\theequation}{S\arabic{equation}}
	\renewcommand{\thefigure}{S\arabic{figure}}
	\renewcommand{\bibnumfmt}[1]{[S#1]}
	\renewcommand{\citenumfont}[1]{S#1}

	\begin{center}
		\textbf{\large Supplementary material for: \\ Autonomous Distribution of Programmable Multiqubit Entanglement in a Dual-Rail Quantum Network}
	\end{center}
	
	
	\section{Full model and derivation of the effective master equation}
	In this section, we present a more detailed discussion of the full model for the cascaded quantum network shown in Fig. 1 of the main text and the derivation of the effective master equation for the qubit state $\rho_{\rm q}$ in the broadband-amplifier limit.

	\subsection{Model} 
	By assuming that the waveguides are sufficiently broadband and by moving into a rotating frame with respect to the photon frequencies $\omega_A$ and $\omega_B$, we can use the standard framework of cascaded quantum systems~\cite{Carmichael93_supp,Gardiner93_supp,QuantumNoise_supp} and model the dynamics of the full quantum network by a master equation of the form ($\hbar$=1)
	\begin{flalign}\label{EqS:ME}
		\dot{\rho}=\left( \mathcal{L}_{\rm ph} +\mathcal{L}^0_{\rm q}+\mathcal{L}_{\rm casc} \right)\rho.
	\end{flalign}
	Here, the first term describes the parametric amplifier with   	
	\begin{flalign}
		\mathcal{L}_{\rm ph}  \rho=-i [H_\chi,\rho] + \sum_{\eta=A,B} \kappa_\eta \mathcal{D}[a_\eta] \rho,
	\end{flalign}	
	where $H_\chi=i g (a_A^\dagger a_B^\dagger-a_Aa_B)$ and $\kappa_\eta$ are the photon decay rates. The Liouville operator describing the bare qubit dynamics reads
	\begin{flalign}
		\mathcal{L}^0_{\rm q}  \rho= \sum_{\eta,i}  \left( -i\frac{\delta_{\eta,i}}{2} [\sigma_{\eta,i}^{z},\rho] + \gamma_{\eta,i} \mathcal{D}[\sigma_{\eta,i}^-] \rho + \frac{\gamma_\phi}{2} \mathcal{D}[\sigma_{\eta,i}^z] \rho \right),
	\end{flalign}
	where $\delta_{\eta,i}=\omega^{\rm q}_{\eta,i}-\omega_\eta$ is the detuning of the qubit frequency $\omega^{\rm q}_{\eta,i}$ from the central photon frequency $\omega_\eta$ and $\gamma_{\eta,i} $ is the qubit decay rate. We have also included a dephasing of each qubit with rate $\gamma_\phi$. Finally, the last term in Eq.~\eqref{EqS:ME} accounts for the cascaded interaction between all the nodes along each waveguide and is given by
	\begin{equation}
		\mathcal{L}_{\rm casc}  \rho= \sum_{\eta,i}\sqrt{\kappa_{\eta}\gamma_{\eta,i}(1-\epsilon^\eta_{0,i}) } \mathcal{T}[a_\eta,\sigma_{\eta,i}^-]\rho +  \sum_{\eta,j>i}  \sqrt{\gamma_{\eta,i} \gamma_{\eta,j} (1-\epsilon^\eta_{i,j})} \mathcal{T}[\sigma^-_{\eta,i},\sigma_{\eta,j}^-]\rho.
	\end{equation}
	Here, we have included the additional parameters $\epsilon^\eta_{i,j}\in [0, 1]$ to model losses in the system. In general, $| \epsilon^\eta_{i,j}|$ is the loss probability for a photon propagating between node $i$ and node $j$, where the index $i=0$ refers to the parametric amplifier.  In this way the $\epsilon^\eta_{i,j}$ can be adjusted to model linear absorption losses, but also parasitic loss channels for the qubits. 
	
	Note that in Eq.~\eqref{EqS:ME} we have absorbed all propagation phases into a redefinition of the qubit operators. This means that up to local phase rotations, all results presented in this work are independent of the precise location of the qubits. This is in contrast to entanglement schemes in bidirectional channels, which 
	typically require specific arrangements~\cite{Agarwal90_supp,GutierrezJauregui23_supp}. Note, however, that the validity of Eq.~\eqref{EqS:ME} assumes that propagation times between the nodes are negligible compared to the relevant timescales of the system dynamics. We will relax this assumption later below.

	\subsection{Adiabatic elimination of the photonic modes} 
	In the limit $\kappa_\eta\rightarrow \infty$ the dynamics of the parametric amplifier modes can be adiabatically eliminated to derive an effective master equation for the reduced density operator $\rho_{\rm q}={\rm Tr}_{\rm ph}\{ \rho\}$ of the qubits only.  To do so, first, we rewrite the full master equation as 
	\begin{flalign}\label{EqS:MEv2}
		\dot{\rho}=\left( \mathcal{L}_{\rm ph}  +\mathcal{L}_{\rm q}+\mathcal{L}_{\rm ph-q} \right)\rho,
	\end{flalign}
	where
	\begin{equation}
		\mathcal{L}_{\rm q} \rho = \mathcal{L}_{\rm q}^0 \rho +   \sum_{\eta,j>i}  \sqrt{\gamma_{\eta,i} \gamma_{\eta,j} (1-\epsilon^\eta_{i,j})} \mathcal{T}[\sigma^-_{\eta,i},\sigma_{\eta,j}^-]\rho
	\end{equation}
	now includes all waveguide-mediated interactions among the qubits, while  
	\begin{equation}\label{EqS:Lphq}
		\mathcal{L}_{\rm ph-q}   \rho= \sum_{\eta=A,B}\sqrt{\kappa\gamma_\eta} \left([a_\eta \rho, \tilde L^\dag_\eta] + [\tilde L_\eta, \rho a_\eta^\dag ]   \right).
	\end{equation}
	Here we have set $\gamma_\eta={\rm max}\{\gamma_{\eta,i}\}$ and introduced the collective operators
	\begin{equation}
		\tilde L_\eta= \sum_i \sqrt{\frac{\gamma_{\eta,i}}{\gamma_\eta} (1-\epsilon^\eta_{0,i})} \sigma_{\eta,i}^-.
	\end{equation}
	In this form, the master equation in Eq.~\eqref{EqS:MEv2} is identical to the master equation of the 2-qubit setup considered in Ref.~\cite{Agusti22_supp}, but with the collective operators $\tilde L_\eta$, instead of $\sigma_\eta^-$, appearing in the photon qubit interaction in Eq.~\eqref{EqS:Lphq}. Therefore, for the adiabatic elimination we can follow the same steps as in Appendix A of Ref.~\cite{Agusti22_supp} and we obtain
	\begin{equation}
		\begin{split}
			\dot{\rho}_{\rm q}(t)
			=   
			\,&\mathcal{L}_{\rm q}\rho_{\rm q}(t)- \sum_{\eta=A,B} \frac{\gamma_\eta}{2}  N^{\eta}_{\rm ph}  \left( [\tilde L_\eta, [\tilde L_\eta^\dag,\rho_{\rm q}]]+  [\tilde L^\dag_\eta, [\tilde L_\eta,\rho_{\rm q}]]\right)\\
			+&\sqrt{\gamma_A\gamma_{B}}  \left( M_{\rm ph}^*[\tilde L_A, [\tilde L_B,\rho_{\rm q}]] +M_{\rm ph}  [\tilde L_A^\dag, [\tilde L_B^\dag,\rho_{\rm q}]]\right),
		\end{split}
	\end{equation}
	where 
	\begin{eqnarray}
		N^{\eta}_{\rm ph}  &=&2 \kappa_\eta  \mathrm{Re}     \int_{0}^{\infty}\mathrm{d}t \, \langle a_\eta^\dag (t)a_\eta(0)\rangle, \\
		M_{\rm ph} &=& \sqrt{\kappa_A\kappa_B}     \int_{0}^{\infty}\mathrm{d}t \, \left( \langle a_A (t)a_B(0)\rangle +\langle a_B (t)a_A(0)\rangle \right). 
	\end{eqnarray}
	For our simple parametric amplifier, we obtain
	\begin{eqnarray}
		N_{\rm ph}=N^\eta_{\rm ph}  &=& g \frac{\sqrt{\kappa_A\kappa_B}}{2}\left[\frac{1}{\left(\frac{\sqrt{\kappa_A\kappa_B}}{2}-g\right)^2}-\frac{1}{\left(\frac{\sqrt{\kappa_A\kappa_B}}{2}+g\right)^2}\right], \\
		M_{\rm ph} &=& g \frac{\sqrt{\kappa_A\kappa_B}}{2}\left[\frac{1}{\left(\frac{\sqrt{\kappa_A\kappa_B}}{2}-g\right)^2}+\frac{1}{\left(\frac{\sqrt{\kappa_A\kappa_B}}{2}+g\right)^2}\right].
	\end{eqnarray}
	Since $M_{\rm ph}^2=N_{\rm ph}(N_{\rm ph}+1)$, we can further express these quantities in terms of the squeezing parameter $r$, i.e., 
	\begin{equation}
		N_{\rm ph}=\sinh^2(r), \qquad  M_{\rm ph}=\sinh(r)\cosh(r), \qquad  \Rightarrow \qquad  r=2\tanh^{-1}(2g/\sqrt{\kappa_A\kappa_B}).
	\end{equation}  
	This relation allows us to write the effective master equation as
	\begin{equation}
		\begin{split}\label{EqS:EffectiveMEgeneral}
			\dot{\rho}_{\rm q}(t)
			=   
			\,&\mathcal{L}_{\rm q}\rho_{\rm q}(t)- \sum_{\eta=A,B} \gamma_\eta \mathcal{D}[\tilde L_\eta]\rho_{\rm q}    + \mathcal{D}[\tilde J_A]\rho_{\rm q}  + \mathcal{D}[\tilde J_B]\rho_{\rm q},
		\end{split}
	\end{equation}
	where
	\begin{flalign}
		\label{Eq:Dissipators}
		&\tilde J_A=\sqrt{\gamma_A} \cosh(r) \tilde L_A- \sqrt{\gamma_B} \sinh(r)\tilde L_B^\dagger,\\
		&\tilde J_B=\sqrt{\gamma_B}\cosh(r)\tilde L_B- \sqrt{\gamma_A}\sinh(r) \tilde L_A^\dagger.
	\end{flalign}
	In the limit of negligible losses, $\epsilon^\eta_{i,j}\rightarrow 0$, we can exploit the fact that for any set of operators $c_i$, the cascaded waveguide interaction,
	\begin{equation}
		\sum_i \mathcal{D}[c_i]\rho + \sum_{j>i}  \mathcal{T}[c_i,c_j]\rho = \mathcal{D}[C]\rho - i [ H_{\rm casc}, \rho],
	\end{equation}
	can be re-expressed in terms of a collective dissipation term with $C=\sum_i c_i$ and a Hamiltonian evolution with 
	\begin{equation}
		H_{\rm casc}= \frac{i}{2}\sum_{j>i}\left( c_i^\dag c_j- c_j^\dag c_i\right).  
	\end{equation} 
	This then leads to 
	\begin{equation}\label{EqS:ReducedME}
		\begin{split}
			\dot{\rho}_{\rm q}(t)
			=   
			\,&-i [H_{\rm casc},\rho_{\rm q}]+ \sum_{\eta,i} \frac{\gamma_\phi}{2} \mathcal{D}[\sigma_{\eta,i}^z] \rho_{\rm q}     + \mathcal{D}[\tilde J_A]\rho_{\rm q}  + \mathcal{D}[\tilde J_B]\rho_{\rm q},
		\end{split}
	\end{equation}
	with
	\begin{equation}
		H_{\rm casc}= \sum_{\eta,i} \frac{\delta_{\eta,i}}{2} \sigma_{\eta,i}^{z}  + \frac{i}{2}\sum_{\eta,j>i} \sqrt{\gamma_{\eta,i}\gamma_{\eta,j}}\left( \sigma_{\eta,i}^+ \sigma_{\eta,j}^- -  \sigma_{\eta,j}^+  \sigma_{\eta,i}^-\right).  
	\end{equation} 
	For identical decay rates, this result reduces to the master equation given in Eq.~(1) in the main text. When general losses are included, we no longer obtain such a simple form, but we can still use Eq.~\eqref{EqS:EffectiveMEgeneral} for numerical simulations. 
	
	\subsection{Propagation times} 
	In the setup, we consider the photons with group velocity $c_\eta$ take the time $\tau_{\eta,i}=d_{\eta,i}/c_\eta$ to propagate the distance $d_{\eta,i}$ between the source and the $i$-th qubit in waveguide $\eta$. In a strict sense, the validity of Eq.~\eqref{EqS:ME} assumes that all these times are negligible compared to the typical timescale of the system evolution. However, due to the cascaded nature of the interaction, the cascaded master equation in Eq.~\eqref{EqS:ME} can still be used to evaluate expectation values for arbitrary large networks. For example, to evaluate steady-state expectation values of the form $\langle O_{\eta,i}O_{\eta',j}\rangle_0$ in the case of non-negligible $\tau_{\eta,i}$, we can use the prescription~\cite{Agusti22_supp}
	\begin{equation}
		\langle O_{\eta,i}O_{\eta',j}\rangle_0 = \langle O_{\eta,i}(\tau_{\eta,i}-\tau_{\eta',j})O_{\eta',j}\rangle|_{\rm Eq.\, (S1)},
	\end{equation}
	where the two-time correlation function on the right-hand side is evaluated with the help of Eq.~\eqref{EqS:ME}. The same applies to the effective qubit master equation in Eq.~\eqref{EqS:ReducedME}, which is only a reduced version of Eq.~\eqref{EqS:ME}. 
	In particular, this means that the qubits can be separated from the source by an arbitrary distance, as long as the relative propagation times satisfy
	\begin{equation}
		\gamma |\tau_{\eta,i}-\tau_{\eta',j}|  \ll 1.
	\end{equation}
	We refer the reader to Ref.~\cite{Agusti22_supp} for a more detailed discussion about the influence of finite time delays on the achievable steady-state entanglement.
	
	\section*{Uniqueness of the steady state}
	\label{Sec:SteadyStateUnique}

	In the main text we have studied the steady states of the effective master equation for $N$ qubit pairs,
	\begin{equation}
		\dot \rho_{\rm q}= \mathcal{L}_N\rho_{\rm q}= -i[H_{\rm casc},\rho_{\rm q}]+  \sum_{\eta=A,B} \gamma \mathcal{D}[J_\eta]\rho_{\rm q}.
	\end{equation}
	It is clear that given a state $|\psi_0\rangle$ that satisfies the dark-state conditions $J_\eta |\psi_0\rangle =0$ and $H_{\rm casc} |\psi_0\rangle= 0$, the density operator $\rho_0=|\psi_0\rangle \langle \psi_0|$ is a pure steady state of this master equation, i.e. $\mathcal{L}_N\rho_0=0$. However, this condition does not guarantee that $\rho_0$ is the unique steady state and there could be other mixed or pure states $\rho_0^\prime$ with $\mathcal{L}_N\rho_0^\prime=0$. The actual steady state of an ideal network would then depend on the precise initial condition, while in practice residual imperfections would create an uncontrolled mixture of multiple possible steady states. This would be detrimental for entanglement generation. 
	
	\subsection*{N=1}
	To prove that the state $|\psi_0(r,\vec \delta_A,P)\rangle$ defined in Eq.~(9) of the main text is indeed the unique steady state of the network for a given detuning pattern  $\vec \delta_A$ and permutation $P$, we start with the case $N=1$ and $\delta_{A,1}=-\delta_{B,1}$. In this case, we can calculate the eigenvalues of the Liouvillian $\mathcal{L}_{N=1}$ analytically and verify that for any finite  $r$ there is only a single eigenvalue $\lambda_0=0$, which corresponds to the state $\rho_0^{(N=1)}=|\Phi_{1,1}^+\rangle\langle \Phi_{1,1}^+|$. We also find that the smallest non-zero eigenvalue is $\lambda_1=\gamma \cosh(2r)/2$ for $r<r^*$ and $\lambda_1=\gamma (6\cosh(2r)-\sqrt{18\cosh(4r)-14})/4$ for $r>r^*$, where $r^*\simeq 0.356$. This eigenvalue determines the gap in the Liouvillian spectrum and, therefore, for any finite $r$ there is a finite relaxation rate toward the steady state.
	
	\subsection{Induction step}
	We now assume that we already know that the product state $\rho_0^{(N)}=|\Phi_\parallel(N)\rangle\langle \Phi_\parallel(N)|$ defined in Eq.~(6) of the main text is the unique steady state of $\mathcal{L}_N$ for $\vec \delta_{A,i}=-\vec\delta_{B,i}$ and that it satisfies the dark-state conditions  $J_\eta |\Phi_\parallel(N)\rangle=0$ and $H_{\rm casc} |\Phi_\parallel(N)\rangle=0$. We now show that under this assumption it is also true that $\rho_0^{(N+1)}$ is the unique steady state of the network with $N+1$ qubit pairs. 
	
	Let us first verify that $|\Phi_\parallel(N+1)\rangle$ is a dark state. The conditions $J_\eta |\Phi_\parallel(N+1)\rangle=0$ are straightforward to verify, since it holds for each qubit pair individually. For the second condition, $H_{\rm casc} |\Phi_\parallel(N+1)\rangle=0$, we write the cascaded Hamiltonian as
	\begin{equation}
		H_{\rm casc}^{(N+1)}= H_{\rm casc}^{(N)}-i \frac{\gamma}{2}\sum_\eta \left(L_\eta(N)\sigma_{\eta,N+1}^+ -L^\dag_\eta(N)\sigma_{\eta,N+1}\right),
	\end{equation}
	and recall that $ |\Phi_\parallel(N+1)\rangle\sim |\Phi_\parallel(N)\rangle\otimes(\cosh(r)|0_{A,N+1}\rangle|0_{B,N+1}\rangle+ \sinh(r)|1_{A,N+1}\rangle|1_{B,N+1}\rangle)$. It follows that 
	\begin{equation}
		\begin{split}
			H_{\rm casc}^{(N+1)}| \Phi_\parallel(N+1)\rangle \sim & \left[ \cosh(r) L_A(N)  - \sinh(r) L^\dag_B(N)\right]  |\Phi_\parallel(N)\rangle \otimes |1_{A,N+1}\rangle|0_{B,N+1}\rangle\\
			&+ \left[ \cosh(r) L_B(N)  - \sinh(r) L^\dag_A(N)\right] |\Phi_\parallel(N)\rangle \otimes |0_{A,N+1}\rangle|1_{B,N+1}\rangle=0.
		\end{split}
	\end{equation}
	
	To prove that $\rho_0^{(N+1)}$ is also the unique steady state, we use the fact that in a fully directional network the reduced steady state of the first  $N$ pairs of qubits, $\rho_0^{(N)}={\rm Tr}_{i=N+1}\{\rho_0^{(N+1)}\}$ is unaffected by adding an additional pair. Further, because $\rho_0^{(N)}$ is pure, there is no entanglement between the subsystems and we can write  $\rho_0^{(N+1)}=\rho_0^{(N)}\otimes \rho^{(x)}_0$, with a steady state $ \rho^{(x)}_0$ for the last pair, which still must be determined (to simplify notation, we use the index $x$ to refer to the extra qubit pair with index $i=N+1$).  To do so we write 
	\begin{equation}
		\begin{split}
			\dot \rho_0^{(N+1)} =& \mathcal{L}_{N}\rho_0^{(N)}  \otimes \rho^{(x)}_0 +\rho_0^{(N)} \otimes  \mathcal{L}_x\rho^{(x)}_0 +  \mathcal{L}_{N-x} \rho_0^{(N+1)}.
		\end{split} 
	\end{equation}
	Here $\mathcal{L}_x\hat{=} \mathcal{L}_{N=1}$ is the single-pair Liouville operator acting on the state of the last qubit pair and 
	\begin{equation}\label{EqS:CrossLiouvillian}
		\begin{split}
			\mathcal{L}_{N-x} \rho_0^{(N+1)} =&-  \frac{\gamma}{2} \sum_{\eta=A,B}  \left[ L_\eta (N) \sigma_{\eta,x}^+ -  L^\dag_\eta (N) \sigma_{\eta,x}^- ,\rho_0^{(N+1)}\right] \\
			&- \frac{\gamma}{2}\left\{ J^\dag _A (N) \left(\cosh(r) \sigma_{A,x}^- -\sinh(r) \sigma_{B,x}^+\right)  + J^\dag _B(N) \left(\cosh(r) \sigma_{B,x}^- -\sinh(r) \sigma_{A,x}^+\right) ,\rho_0^{(N+1)}  \right\}_+
		\end{split} 
	\end{equation}
	accounts for the remaining cross terms. Note that here we have already used that $J_\eta(N) \rho_0^{(N+1)} = \rho_0^{(N+1)}J_\eta^\dag=0$ due to the dark-state condition for $|\Phi_\parallel(N)\rangle$. By looking at all the different contributions in Eq.~\eqref{EqS:CrossLiouvillian} we can collect terms such as
	\begin{equation}
		\begin{split}
			\left[ - L_A(N) + \sinh(r) J_B^\dag(N)  \right] \sigma_{A,x}^+  \rho_0^{(N+1)} = &\left[ - L_A(N) + \sinh(r)\cosh(r) L_B^\dag(N) - \sinh^2(r) L_A(N)  \right] \sigma_{A,x}^+  \rho_0^{(N+1)}\\
			=& -\cosh(r) \underbrace{\left[ \cosh(r) L_A(N) - \sinh(r) L_B^\dag(N)  \right]}_{=J_A(N)} \sigma_{A,x}^+  \rho_0^{(N+1)} =0,
		\end{split}
	\end{equation} 
	and find that they vanish independently of $\rho^{(x)}_0$. The same is true for other combinations such that $\mathcal{L}_{N-x} \rho_0^{(N+1)}=0$. Therefore, when tracing over the first $N$ qubit pairs, the steady state $\rho_0^{(x)}$ satisfies
	\begin{equation}
		\rho_0^{(x)}= \mathcal{L}_x\rho_0^{(x)} =0,
	\end{equation}
	which has a unique solution given by $\rho_0^{(x)}=|\Phi^+_x\rangle\langle \Phi^+_x|$. 
	
	\subsection*{General detunings} 
	Finally, we consider  non-trivial detuning patterns $\vec \delta_B=-P \vec \delta_A$ and show that also in this case the steady state $\rho^{(N)}_0=|\psi_0(r,\vec \delta_A,P)\rangle\langle \psi_0(r,\vec \delta_A,P)|$ is unique. This can be done by simply assuming that there is another steady state $\rho_0^\prime\neq \rho^{(N)}_0 $ of the Liouvillian $\mathcal{L}_N$. Then we can simply invert the arguments about the form invariance of the master equation presented in the main text below Eq.~(8) and obtain a steady state for the network with $\vec \delta_B=-\vec \delta_A$,
	\begin{equation}
		\rho_0^\prime(\vec \delta_B=-\vec \delta_A)=  \mathcal{U}^\dag \rho_0^\prime \mathcal{U},\qquad  \mathcal{U}= \prod_\sigma U_{i_\sigma,i_\sigma+1}.
	\end{equation}  
	However, since we know that there is only one unique steady state for this detuning pattern, it means that $\rho_0^\prime(\vec \delta_B=-\vec \delta_A)= |\Phi_\parallel(N)\rangle\langle \Phi_\parallel(N)|$ and $\rho_0^\prime=|\psi_0(r,\vec \delta_A,P)\rangle\langle \psi_0(r,\vec \delta_A,P)|$.

	\section*{Scaling of entanglement in realistic networks}
	In the main text we have shown that in a realistic setup, the entanglement along the chain is degraded by finite-bandwidth effects and qubit dephasing. In this section, we provide some additional numerical results on the dependence of $N_{\rm ent}$ on the squeezing strength and on the influence of waveguide losses and imperfect chiral couplings.  
	
	\subsection*{Squeezing strength}
	
\begin{figure}
		\centering		
		\includegraphics[width=0.9\columnwidth]{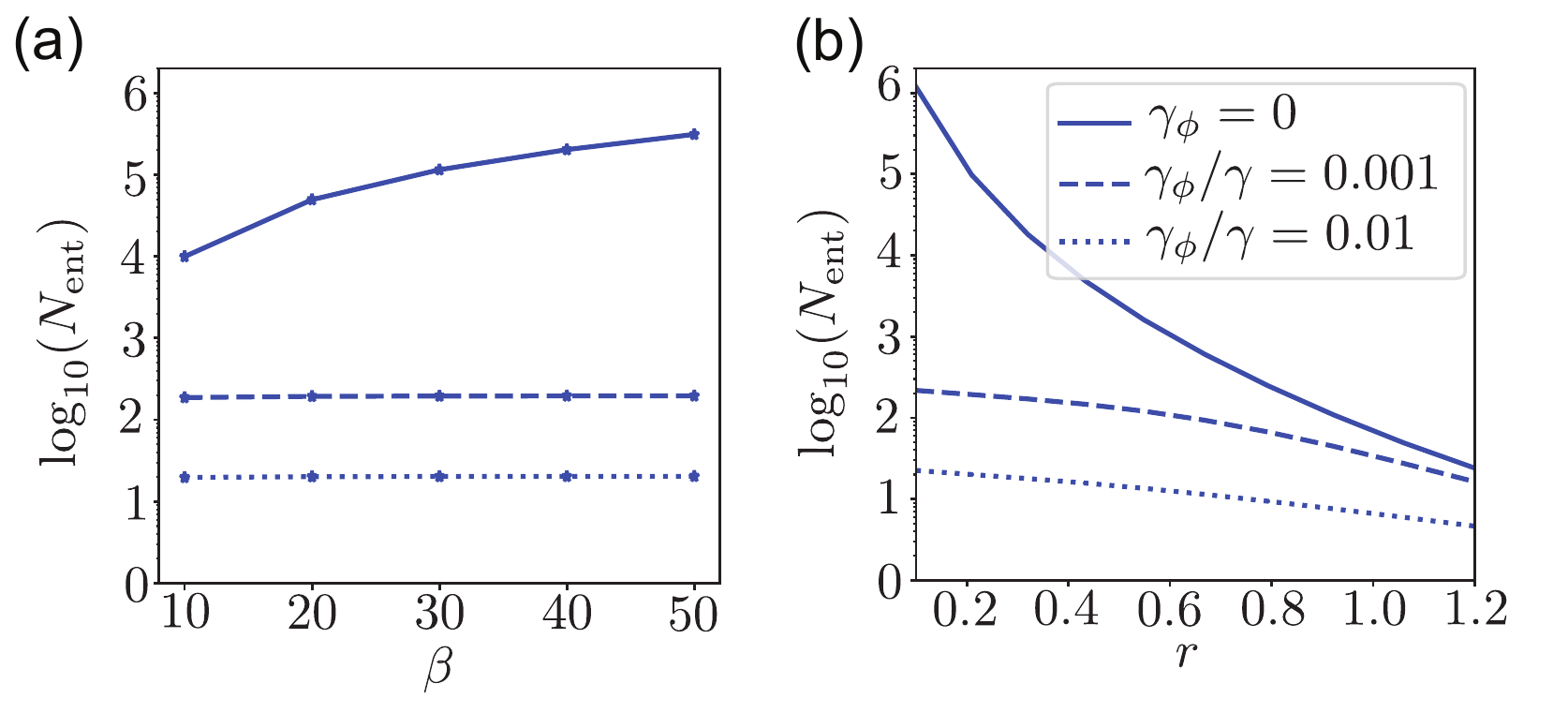}
		\caption{Maximal number of entangled pairs, $N_{\rm ent}$, as a function of (a) $\beta=\kappa/\gamma$ for different dephasing rates $\gamma_{\phi}$ at fixed squeezing strength $r=0.2$ and (b) as a function of the squeezing strength $r$ for the same dephasing rates $\gamma_\phi$ at fixed $\beta=30$. For both plots, we have assumed $\vec{\delta}_A$=0.}
		\label{fig:FigS1Squeezing}
	\end{figure}

	In Fig.~\ref{fig:FigS1Squeezing} (a), we plot $N_{\rm ent}$ as a function of $\beta=\kappa/\gamma$ for a small squeezing parameter of $r=0.2$. Compared to Fig. 4(b) in the main text, the decay of entanglement along the chain is much slower and also more robust  with respect to qubit dephasing. Note, however, that the maximal concurrence of $\mathcal{C}_{11}\approx 0.38$ is also much lower than for $r=1$. In Fig.~\ref{fig:FigS1Squeezing} (b) we fix the value of $\beta=30$ and plot $N_{\rm ent}$ as a function of $r$. This plot shows the expected trade-off between a high degree of entanglement and the number of entangled pairs. 
	
	\begin{figure}
		\centering		
		\includegraphics[width=0.9\columnwidth]{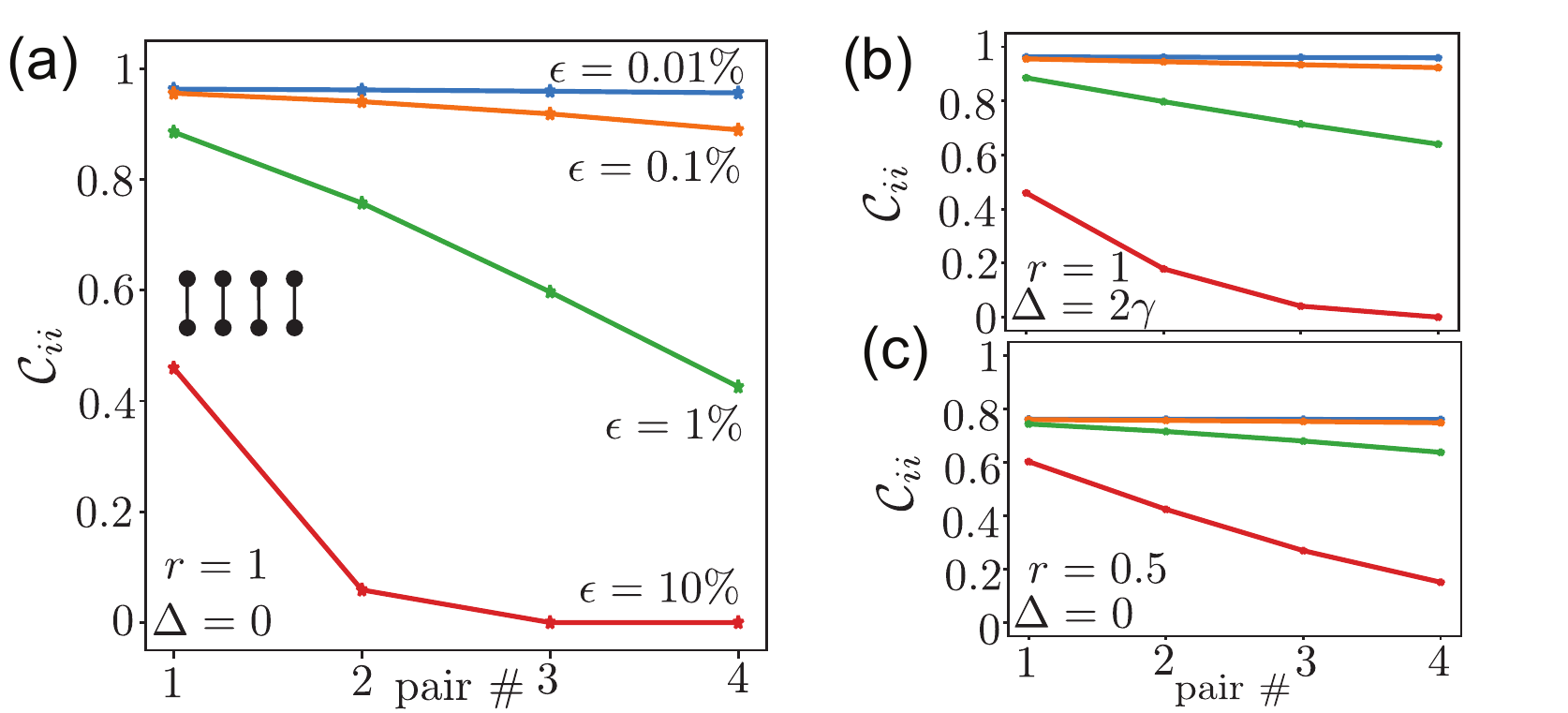}
		\caption{(a) Effect of finite propagation losses $\epsilon>0$ on the steady-state entanglement for $N=4$ for (a) squeezing strength $r=1$ and on resonance $\Delta=0$, (b) for $r=1$ and parallel configuration with detuning $\Delta=2\gamma$, and (c) for smaller squeezing strength $r=0.5$ and on resonance $\Delta=0$. For a small value of $\epsilon$, this plot predicts an approximately linear decay of the entanglement along the waveguide. For all plots, $\epsilon^\eta_{i,j}=\epsilon |i-j|$ and $\beta\rightarrow \infty$ have been assumed.}
		\label{fig:FigS2Losses}
	\end{figure}
	
	\subsection*{Waveguide losses}

	In Fig.~\ref{fig:FigS2Losses}, we consider a chain of $N=5$ pairs of qubits and simulate the effect of losses along the waveguides. For concreteness, we assume $\epsilon^\eta_{i,j}=\epsilon (i-j)$, which means that there is a fixed loss probability $|\epsilon|$ between two successive nodes of the network.

	\subsection*{Nonideal chiral coupling}
	
	Throughout the main text, we have considered the waveguide to be completely directional, that is, all the photons propagate along the same direction. While this allows us to get analytical results, realizing such directional interactions will only be possible with a certain fidelity. Here we present additional numerical results for waveguides, the qubits along the waveguides can decay into left-propagating modes with rate $\gamma_L$ and into right-propagating modes with rate $\gamma_R$. The main results are then recovered when $\gamma_L=0$ and $\gamma_R=\gamma$.
	
	To extend our model to a bi-directional waveguide, we include the effect of an additional left propagating channel into our effective qubit master equation (see, e.g., Ref.~\cite{Pichler15_supp}). We obtain 
	\begin{equation}\label{eq:ME_bidirectional}
		\dot{\rho}_{\rm q}=-i[H_{\rm chiral},\rho_{\rm q}]+\sum_{\eta=A,B} \gamma_{R}\mathcal{D}[J_\eta]\rho_{\rm q}+\sum_{\eta=A,B} \gamma_{L}\mathcal{D}[L_\eta]\rho_{\rm q}.
	\end{equation}
	The new modes contribute to both the coherent and incoherent interaction. The coherent interaction now depends on the difference between left- and right-modes and vanishes at a completely bi-directional waveguide,
	\begin{equation}
		H_{\rm chiral}=\frac{i(\gamma_R-\gamma_L)}{2}\sum_{\eta,j>i}(\sigma^+_{\eta,i}\sigma^-_{\eta,j}-{\rm H.c}).
	\end{equation}
	For the incoherent term, we have assumed that the left-propagating modes decay into vacuum modes with a collective jump operator $L_{\eta}=\sum_{i=1}^N\sigma_{\eta,i}$. Therefore, only the right-propagating modes are squeezed and correlated. Note that this form also assumes that the qubits are spaced by multiples of the central wavelength, such that all propagation phases cancel.
	
\begin{figure}
		\centering		
		\includegraphics[width=0.9\columnwidth]{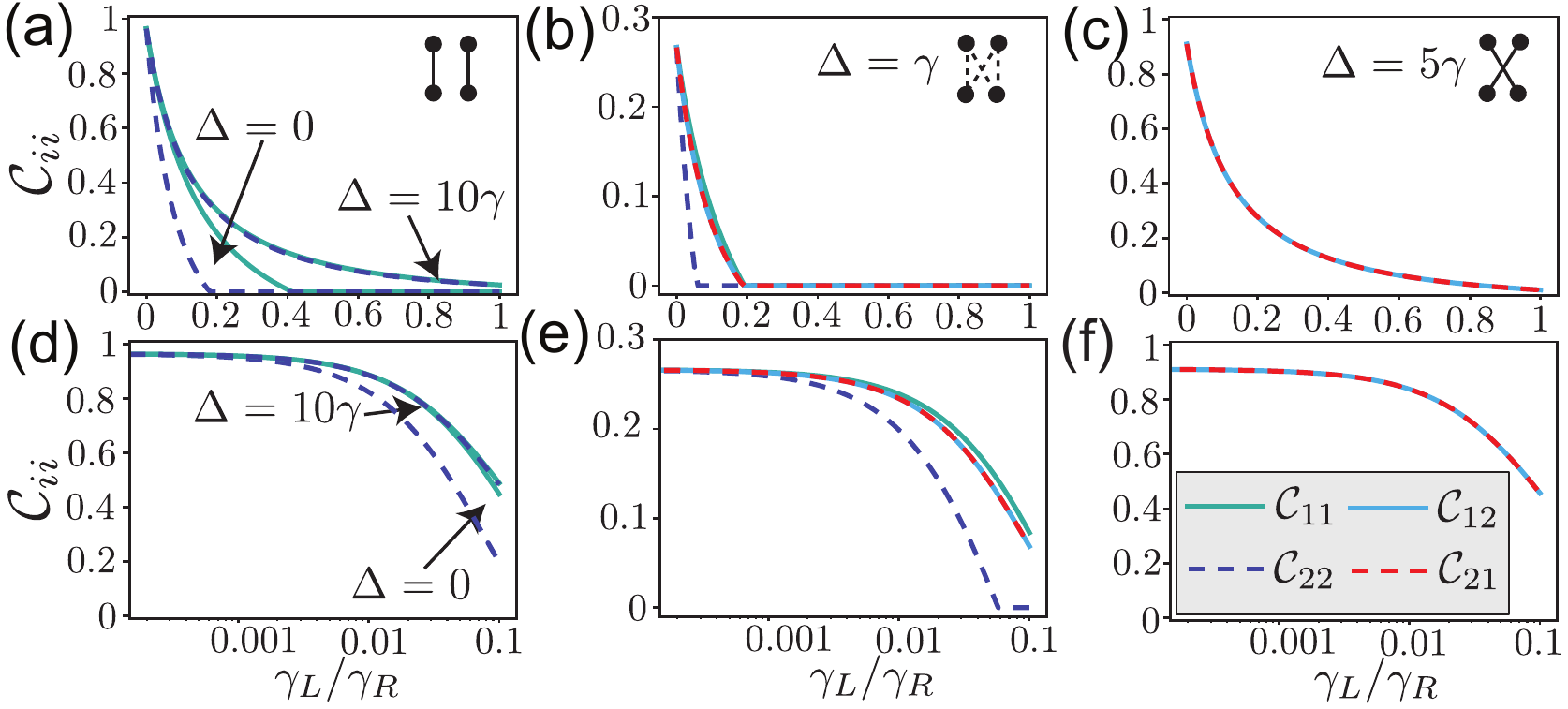}
		\caption{Plot of the bipartite concurrence of the steady state of the master equation in Eq.~\eqref{eq:ME_bidirectional}, which includes a decay into left-propagating waveguide modes with rate $\gamma_L$. The plots in (a) and (d) assume parallel detunings with $\vec{\delta_B}=-\vec{\delta_A}$ and $\delta_{A,i}=\Delta(i-1)$, while the plots in (b), (c), (e) and (f) assume a reversed detuning pattern $\vec{\delta_B}=-P_{\rm rev} \vec{\delta_A}$. In all plots $r=1$.}
		\label{fig:FigS3Chiral}
	\end{figure}

	In Fig.~\ref*{fig:FigS3Chiral}, we numerically solve this master equation for different degrees of chirality, $\gamma_L/\gamma_R$. We observe that the effect of a finite $\gamma_L$ on the resulting steady state depends a lot on the type of entanglement, which in turn is determined by the detunings. While multipartite entangled states are strongly affected by a finite bi-directional coupling, bipartite entangled states are more robust, and in the far-detuned regime, a finite amount of entanglement survives up to  $\gamma_L/\gamma_R=1$.
	
	\subsection*{Example: Microwave quantum networks}
	
	To illustrate the performance of this protocol in a realistic setup, where all types of imperfections are taken into account, we consider in this subsection the example of superconducting qubits connected via microwave transmission lines.
	
	As a starting point, we use the parameters from a recent work by Joshi \textit{et al.}~\cite{Mirhosseini22_supp}, which describes the realization of a chiral coupling of a superconducting qubit to a microwave waveguide. From this reference, we deduce a qubit dephasing rate of $\gamma_{\phi}/2\pi= 50$ kHz (in accordance with other state of the art experiments~\cite{Kannan22_supp} ), a directional emission rate of $\gamma_R/2\pi\simeq 1$ MHz and an unwanted decay into the opposite direction with rate $\gamma_L/\gamma_R\simeq 0.01$. In addition, in this experiment, there is a residual decay into non-guided modes with a rate $\gamma^\prime/2\pi= 364$ kHz. In our numerical simulations, we include this process by adding a new term $\gamma^\prime\sum_{\eta,i}\mathcal{D}[\sigma_{\eta,i}]\rho$ to our master equation. For these parameters, $\Delta=0$ and assuming an ideal two-mode squeezing source with $r=1$, we obtain $N_{\rm ent}\simeq 1$ and the concurrence of the first pair is $\mathcal{C}_{11}\simeq 0.1$. Obviously, this poor result is mainly related to the large residual decay rate $\gamma'$. By assuming that this decay channel can be eliminated in future setups, $\gamma'\rightarrow 0$~\cite{Mlynek14_supp}, the result improves to $N_{\rm ent}\simeq 2$
	and $\mathcal{C}_{11}\simeq 0.53$, now being primarily limited by decoherence with rate $\gamma_\phi/\gamma\simeq 0.05$.  
	
	Let us now consider the same parameters, but assuming the finite detunings $\delta_A=(i-1)\Delta$ with  $\Delta=\gamma_R$. Consistent with Fig. 4 (d) in the main text, we find that while keeping $\mathcal{C}_{11}\simeq 0.53$, this modification would already boost the total number of entangled pairs to about $N_{\rm ent}\simeq 20$ (assuming $\gamma^\prime=0$). Further, by improving the ratio $\gamma_\phi/\gamma_R$ by a factor of ten (which is well within the range of typical qubit coherence times) would boost this number to about $N_{\rm ent}\simeq 120$ (with $\Delta=\gamma_R$) and $N_{\rm ent}\simeq 6$  (with $\Delta=0$) and with $C_{11}\simeq 0.85$.  At this stage the effects of a finite rate $\gamma_L$ become relevant.

	Let us now address the effect of a finite amplifier bandwidth. In the microwave regime, two-mode squeezing sources are usually realized with Josephson parametric amplifiers (JPAs) or travelling wave parametric amplifiers (TWPAs). Typical bandwidths for these devices are  $\kappa_{\rm JPA}\simeq 2\pi\times 10$ MHz~\cite{Fedorov21_supp} and $\kappa_{\rm TWPA}\simeq 2\pi\times 1$ GHz~\cite{Esposito21_supp,Esposito22_supp}. Combining the parameters from above with the JPA, the relevant ratio between the amplifier bandwidth and the qubit decay rate is $\beta_{\rm JPA}=\kappa_{\rm JPA}/\gamma_R\simeq 10$. In this case, the finite bandwidth does not change the conclusion from above for $\Delta=0$ and we obtain $N_{\rm ent}\simeq 2$ with $C_{11}=0.45$ for $\gamma^\prime=0$. For $\Delta=\gamma_R$ our extrapolation predicts $N_{\rm ent}\approx 50$, but since we must limit the maximum detuning to $\delta_{\rm max}<\kappa$, the limit in this example is set by $N_{\rm ent}\approx \beta \simeq 10$.
	
	To go beyond this limit, we can use a TWPA. In this case, the bandwidth ratio can reach values up to $\beta_{\rm TWPA}=\kappa_{\rm TWPA}/\gamma_R\simeq 10^3$ and all the results for $N_{\rm ent}$ and $\mathcal{C}_{11}$ reduce to the infinite-bandwidth results from above. Note, however, that this assumes an ideal amplifier without any added noise. 
	
	In summary, these estimates show that while the preparation of highly entangled multi-qubit states naturally requires sophisticated experimental setups, existing experimental techniques in the field of superconducting circuits are in principle already enough to demonstrate the simultaneous entanglement of $N\approx2-10$  qubit pairs or generate multipartite entanglement among $\sim4-8$ separated qubits.

\end{document}